\documentstyle[12pt,fleqn,epsfig]{article}

\begin{document}
\begin{center}
{\Huge Shaping Robust System through Evolution}
\end{center}

\begin{Large}
\begin{center}
Kunihiko Kaneko\footnotemark[1]\footnotemark[2]{}
\end{center}
\end{Large}
\begin{center}
\noindent
\footnotemark[1] Department of Basic Science, Univ. of
Tokyo, 3-8-1 Komaba, Tokyo 153-8902, Japan\\
\footnotemark[2] ERATO Complex Systems Biology Project, JST, 3-8-1
Komaba, Tokyo 153-8902, Japan \\
\end{center}

{\Large \bf Abstract}

Biological functions are generated as a result of developmental dynamics that
form phenotypes governed by genotypes. The dynamical  system for development  is shaped through genetic evolution 
following natural selection based on the fitness of the phenotype. Here we study how this dynamical system is
robust to noise during development and to genetic change by mutation.
We adopt a simplified transcription regulation network model to govern gene expression, which gives a fitness function. 
Through simulations of the network that undergoes mutation and
selection, we show that a certain level of noise in gene expression is required
for the network to acquire both types of robustness.
The results reveal how the noise that
cells encounter during development shapes any network's robustness, 
not only to noise but also to
mutations. We also establish a relationship between developmental and mutational robustness
through phenotypic variances caused by genetic variation and epigenetic noise.
A universal relationship between the two variances is derived, akin to the fluctuation-dissipation
relationship known in physics.

{\Large \bf Lead paragraph}

{\bf
Biological function is generally robust to noise in dynamical systems
and to mutation to structure of the system.
How are these two types of robustness related?
Through numerical simulations of a minimal model that captures the essence of a gene expression
dynamics that undergoes a mutation and selection process, we demonstrate that the
evolved networks acquire robustness to mutation only when gene expression is
sufficiently noisy - a physical constraint highlighted by a series of experimental
studies in recent years.  We derive a quantitative condition for evolution of
robustness which ultimately links  robustness to mutation in the evolutionary
time-scale and robustness to noise in development in the reproductive time-scale.
Our model simulations predict correlation between the two robustness,
which lead to a universal proportionality relationship between variances of genetic and epigenetic
phenotype fluctuations, which remind us of fluctuation-dissipation relationship in statistical
physics.
}

\section{Introduction}

Function in a biological system is generated by dynamical systems in general.
A specific shape of a protein that gives rise to some function is formed by folding dynamics. 
In a cell, gene expression pattern changes in time, to shape some chemical composition, from which
a given function of the cell emerges. In multicellular organisms, developmental dynamics lead to patterns of differentiated cells.
Phenotype is shaped in a broad sense from such `developmental dynamical systems',
according to which the fitness of an individual is determined.

In a biological system, such developmental dynamics are shaped through Darwinian evolution. Offspring are produced
depending on the fitness of the parents, with some variations that introduce slight modifications in parameters or networks in
developmental dynamical systems. Among modified dynamical systems, phenotypes with higher function are selected 
for the next generation. Hence, evolutionary shaping of function is a result of variation and selection of developmental dynamical systems.

Combining both the terminology of biology and dynamical systems,
evolutionary process is summarized as follows. 
(i) The genotype gives a set of equations, i.e., terms and parameters in developmental dynamical systems.
(ii) From an initial condition, the developmental dynamical system leads to a set of state variables for a phenotype.
(iii) Fitness is a function of these state variables. The offspring number increases with fitness.
(iv) In reproduction, there are mutations (or other sources for variations) giving rise to variation in dynamical systems.
At each generation, there is redistribution of dynamical systems. If evolution serves to increase fitness, dynamical systems 
with higher function are shaped through this selection-mutation process.

In discussing this evolutionary shaping of dynamical systems with higher function, an important issue is robustness.
Robustness is defined as the ability to function against changes in the parameters 
of a given system \cite{Evolution,Wagner,Barkai-Leibler,Robustness-Alon,Plos1,Wagner2}. In any biological system, these changes have two distinct origins: 
genetic and epigenetic. The former concerns structural robustness of the phenotype, i.e., rigidity of phenotype against variation
in dynamical systems, introduced by genetic changes produced by mutations. 
On the other hand, the latter concerns robustness against the stochasticity that can arise in a given dynamical system, which
includes fluctuation in initial states and stochasticity occurring during developmental dynamics or in the external environment. 
For example, stochasticity in gene expression has recently been studied extensively both experimentally 
and theoretically \cite{Elowitz,Collins,Furusawa,Barkai,noise-review}. Indeed, the existence of such stochasticity is
natural, because the number of molecules in a cell is generally limited.
Accordingly, phenotypes of cells differ, even among those sharing the same genotype\cite{Koshland}.
An important recent recognition is that phenotypic noise is indeed significant, as is highlighted by the log-normal distribution of protein abundances in bacteria\cite{Furusawa}.

The two types of robustness---epigenetic and genetic---are concerned with 
fluctuations in phenotype caused by noise in developmental processes and by
variations in dynamical systems caused by genetic changes.
In terms of dynamical systems, these two types of robustness are the stability of a state (an attractor) to external noise and 
the structural stability of the state against changes in the underlying equations, respectively.
Then, how are these two forms of robustness related in evolution? 
Recently we have found a possible relationship between the two, from an experiment on adaptive evolution
in bacteria\cite{Sato} and numerical evolution of reaction networks and transcription regulation networks\cite{Plos1,JTB}.
The two types of robustness, measured by the (inverse of) phenotypic variances,
increase in correlation through evolution, whereas evolutionary speed is greater
as the magnitude of the phenotypic fluctuation due to noise is increased.
Hence, the relevance of noise to evolution is suggested.

Despite recent quantitative observations on phenotypic fluctuation,
noise is often thought to be an obstacle in tuning a system to achieve and maintain a state with higher functions, 
because the phenotype may be deviated from an optimal state achieving a higher function through such noise. 
Indeed, the question most often 
asked is how a given biological function can be maintained in spite of the existence of phenotypic noise\cite{noise-review,Ueda}.
Although the positive roles of stochasticity in gene expression to cell differentiation \cite{IDT}
and adaptation have been discussed \cite{Kashiwagi,CFKK_PLoS}, its role in evolution has not been explored fully.
As a relatively large amount of phenotypic noise has been preserved through 
evolution, it is important to investigate any positive roles of such noise for the evolution of
biological functions.

The present paper is organized as follows. Following the Ref.\cite{Plos1}, we introduce a simple
transcription regulation network model in \S 2.  In this model. fitness is determined by the gene expression pattern
generated by this network. Using a genetic algorithm to change the network so that the
fitness is increased, we show that 
evolution of the robustness of fitness to mutation to a given network is possible
only under a certain level of noise in the gene expression dynamics. Dependence of
the fitness distribution on the noise level is analyzed in \S 3.
Variances of fitness due to  developmental noise and to genetic variation are computed, which
give indices for developmental and mutational robustness, respectively.
The  proportional relationship between the two variances is given in \S 4, whereas
in \S 5, the two variances are computed for the expression of all genes, which demonstrate common proportion
relationship, suggesting the existence of a universal relationship akin to
fluctuation-dissipation theorem in statistical physics.
Summary and discussion are given in \S 6.

\section{Model}

To study evolutionary shaping of dynamical systems, we consider a simple model for `development'. This
consists of a complex dynamic process to reach a target phenotype under a condition of noise that might
alter the final phenotypic state. We do not choose a biologically realistic model that describes
a specific developmental process, but instead take a model as simple as possible to satisfy
the minimal requirement for our study. Here we take a simplified model for
gene expression dynamics, governed by transcription regulatory networks.
Expression of a gene activates or inhibits expression of other genes under noise.
These interactions between genes are determined by the transcription regulatory network (TRN). The expression profile
changes in time, and eventually reaches a stationary
pattern. This gene expression pattern determines fitness.

To be specific, a typical switch-like dynamics with a sigmoid input--output behavior
\cite{gene-net,Mjolsness,Sole}
was adopted, although several simulations in the form of biological networks
will give essentially the same result. In this simplified model, the dynamics of a given gene expression
level $x_i$ is described by:

\begin{equation}
dx_i/dt =\tanh[\frac{\beta}{\sqrt{M-k}} \sum_{j >k}^{M} J_{ij}x_j]-x_i +\sigma \eta _i(t),
\end{equation}

\noindent
where $J_{ij}=-1,1,0$, and $\eta_i(t)$ is Gaussian white noise given by
$<\eta _i(t)\eta _j(t')>= \delta _{i,j}\delta(t-t')$.  $M$ is the total number of genes, and $k$ is
the number of output genes that are responsible for fitness to be determined.
The value of $\sigma$ represents noise strength
that determines stochasticity in gene expression.
By following a sigmoid function $tanh$, $x_i$ has
a tendency to approach either 1 or -1, which is regarded as `on' or `off' in terms of
gene expression.
The initial condition is given by (-1,-1,...,-1); i.e., all genes are off 
unless noted otherwise.

Depending on the expression pattern, the cell state changes. Hence, the function of the cell and accordingly its fitness changes.
Here we assume that fitness is determined by setting a target gene expression pattern. In the 
present paper we adopt a simple target so that the gene expression levels ($x_i$) for the output genes $i=1,2,\cdots,k \leq M$ reach `on' states, i.e., $x_i>0$. The fitness $F$ is at its maximum if all $k$ genes are `on' after a transient time span $T_{ini}$, and at its minimum if
all are off. $F$ is set at $0$, if all the target genes are on, and is decreased by 1 if one of
the $k$ genes is `off'.
Accordingly, the fitness function is defined by

\begin{equation}
F= \frac{1}{2(T_f-T_{ini})}\sum_{j=1}^k \int_{T_{ini}}^{T_f} (Sign(x_j)-1) dt,
\end{equation}
\noindent
where $Sign(x)=1$ for $x>0$ and $-1$ otherwise.
The initial time $T_{ini}$ can be considered as the time required for developmental dynamics.  
The fitness is computed only after time $T_{ini}$, which is sufficiently large for a given gene expression's dynamics to
fall on an attractor.  In the simulations presented here we adopt $T_{ini}=50$ 
and the results to be discussed are not altered even if it is increased.  
Here, in considering also the possibility of an oscillatory attractor, we adopt the temporal average for the fitness, 
but indeed, it is not necessary as attractors after evolution are mostly fixed points. 
$T_f$ is set at $T_{ini}+50$ here, but even if  it is  defined just by a snapshot value at $t=T_{ini}$,
the results are not essentially altered.

Selection is applied after the introduction of mutation at each generation in the TRN.
Among the mutated networks, we select those with higher fitness values. 
Because the network is governed by $J_{ij}$ which determines the `rule' of the dynamics,
it is natural to treat $J_{ij}$ as a measure of genotype.
Individuals with different genotype have a different set of $J_{ij}$. 

As model (1) contains a noise term, the fitness can fluctuate at each run, which leads to a distribution in $F$,
even among individuals sharing the same network. 
For each individual network, we compute the average fitness $\overline{F}$ over a given number of runs.
At each generation there are $N$ individuals with different sets of $J_{ij}$. 
Then we select the top $N_s (<N)$ networks that have higher fitness values.

At each generation there are $N$ individuals. We compute the average fitness $\overline{F}$ for each network
by carrying out $L$ runs for each. Then, $N_s=N/4$
networks with higher values of $\overline{F}$ are selected for the
next generation, from which $J_{ij}$ is `mutated', i.e.,
$J_{ij}$ for a certain pair $i,j$ selected randomly with a certain fraction is changed
among $\pm1,0$. (To be specific, only the changes $1 \leftrightarrow 0 \leftrightarrow -1$ are allowed). The fraction of path change
is given by the mutation rate $\mu$. 
Unless otherwise mentioned, only a single path, i.e., a single pair of $i,j$
is changed so that the mutation rate $\mu=1/M(M-k)$. Here we make $N/N_s$ mutants from each of the top $N_s$ networks, 
to keep $N$ networks again for the next generation.
Following mutation, the $N$ individuals at each generation have slightly different network elements, 
$J_{ij}$, so that the values of $\overline{F}$ differ.
From this population of networks we repeat the processes of developmental dynamics, their mutation,
and selection of networks with higher fitness values. 

Unless otherwise mentioned, we chose
$N=L=200$, while the conclusion to be shown does not change as long as
these are sufficiently large. (We have also carried out the selection
process by $F$ instead of $\overline{F}$, but the conclusion is not
altered if $N$ is chosen to be sufficiently large.) Throughout the
paper, we use $\beta=50$.
For most numerical results we use $M=64$ and
$k=8$, unless otherwise mentioned. 
Initially we chose $J_{ij}$ randomly with equal probability for $\pm1,0$.
However, the results to be discussed are not changed qualitatively, even if the
fraction of 0 is increased.
As shown in eq.(1), we did not include a connection from the output genes
$i \leq k$, so that $J_{ij}$ with $j\leq k$ is fixed at 0 and is not mutated.
However, the results to be discussed are not altered, even if such connections are included in eq.(1).

\section{Decrease in the average fitness with the {\sl decrease} of noise}

Let us first see how the evolutionary process changes as a function of
the noise strength $\sigma$. Within a hundred generations, the top fitness among the network
population approaches the highest value $0$, if noise level is not too high ($\sigma<  \sigma_c' \approx.12$ for $M=64$) (see Fig. 1).
However, this is not the case for the average fitness among a population with slightly different genes.
For low noise level, average fitness stays lower than the fittest value. In the middle range of noise level, the
average value approaches the fittest value. This difference against the noise level is
clearer in the temporal evolution of the lowest fitness among the existing networks.
As shown in Fig. 1, the value stays rather low for a small noise case, whereas for 
evolution under a high noise value, it goes up with the top fitness value.
In other words, the distribution of the fitness $P(\overline{F})$ over existing networks 
has three distinct behaviors, depending on the noise strength $\sigma$. 
  
\begin{figure}[tbp]
\begin{center}
\includegraphics[width=8cm,height=6cm]{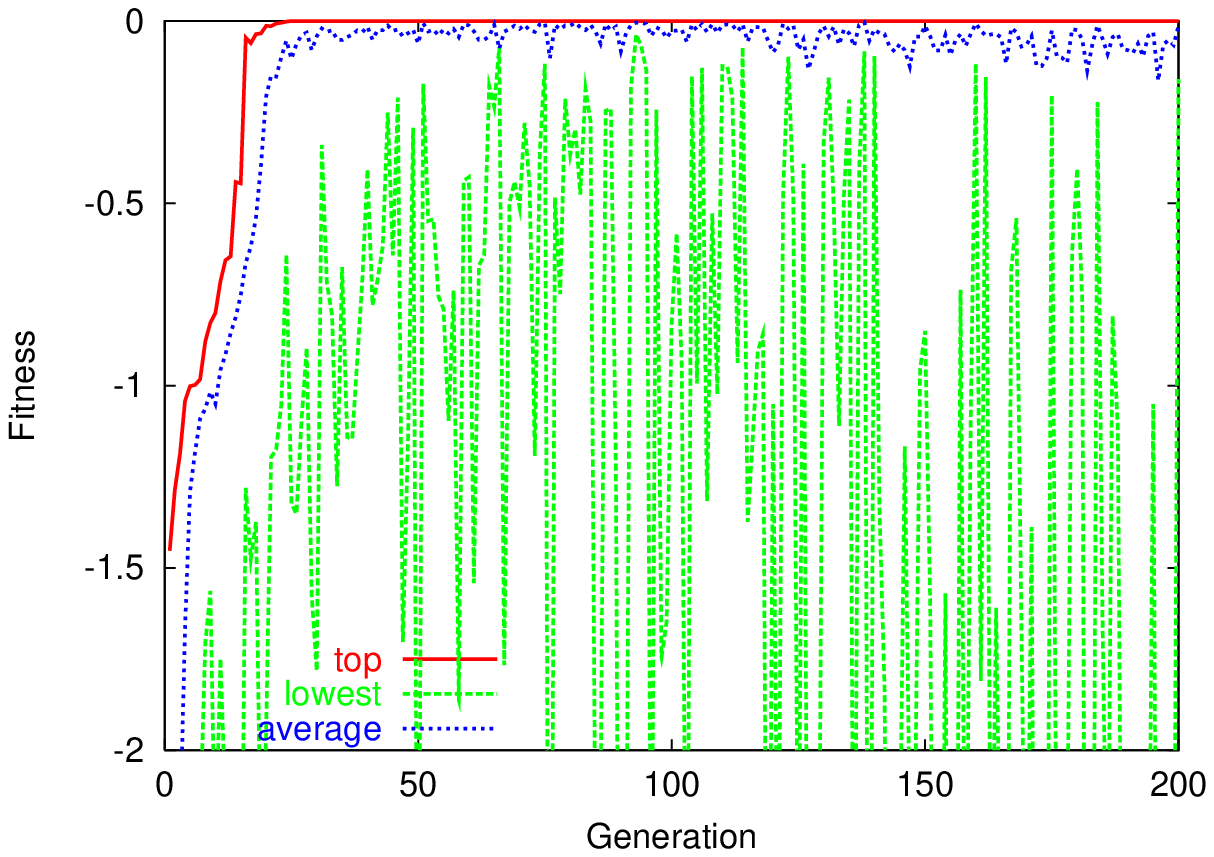}
\includegraphics[width=8cm,height=6cm]{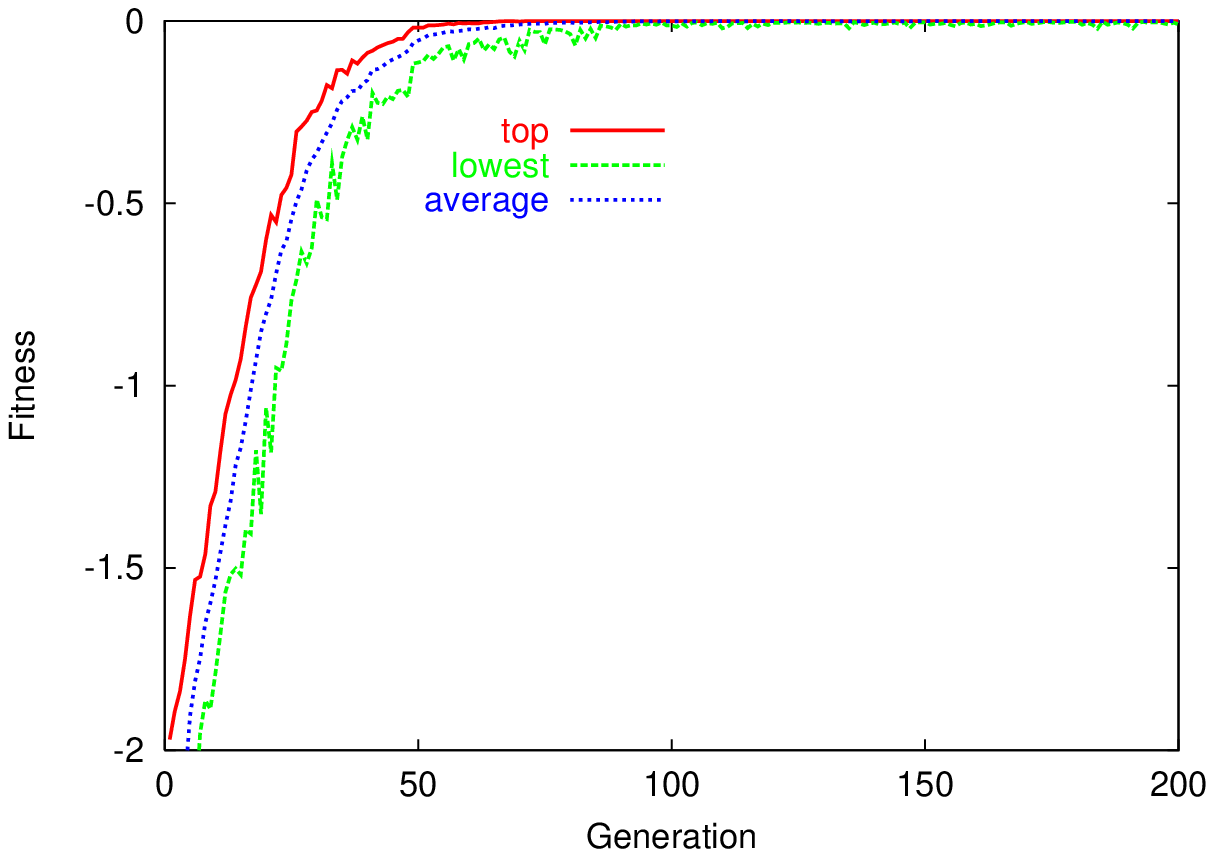}
\includegraphics[width=8cm,height=6cm]{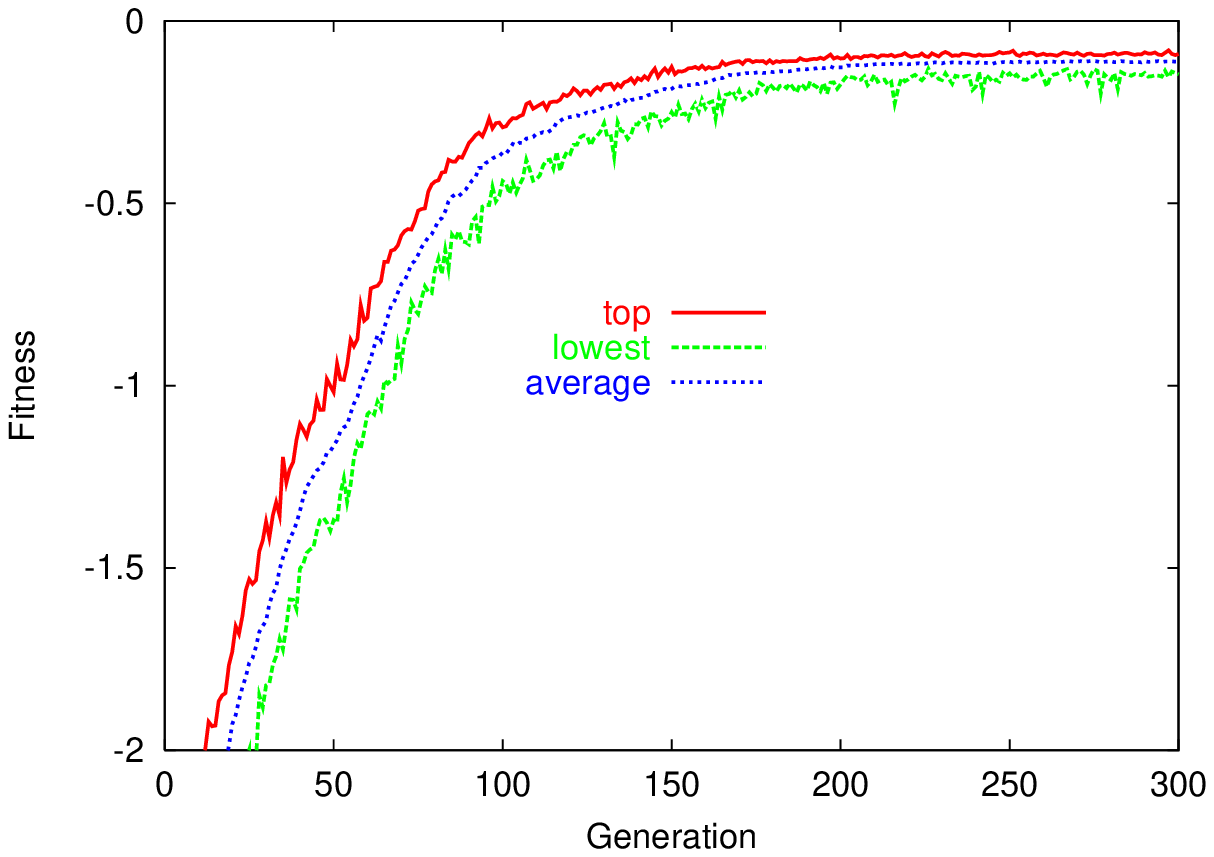}
\caption{
Evolutionary time course of the fitness $\overline{F}$. The highest, average,
and lowest values of the fitness $\overline{F}$ among all
individuals that have different genotypes (i. e., networks $J_{ij}$) 
at each generation are plotted. (a) $\sigma=0.01$, (b) $\sigma=0.1$ (c)$\sigma=0.2$.
}
\end{center}
\end{figure}

(i) For small $\sigma$ ($<\sigma_c \approx 0.04$ for $M=64$ and the mutation rate with 1 path per generation),
the distribution is broad. The top reaches the fittest, although there remain individuals with
very low fitness values $\overline{F}$, even after many generations of evolution.

(ii) For the middle range $\sigma$ ($\sigma_c <\sigma < \sigma_c' $),
the distribution is sharp and concentrated at the fittest value.
Even those individuals with the lowest fitness approach $\overline{F}=0$ .
We call this range the `robust evolution region'.

(iii) For the much larger noise value ($\sigma> \sigma_c'$),
the top does not reach the fittest values, while the distribution is sharp, and concentrated
around the top value.  Here the noise is so large that $x_i$ changes its sign at a certain fraction
even after reached the target, so that the top fittest is not achieved.

\begin{figure}[tbp]
\begin{center}
\includegraphics[width=8cm,height=6cm]{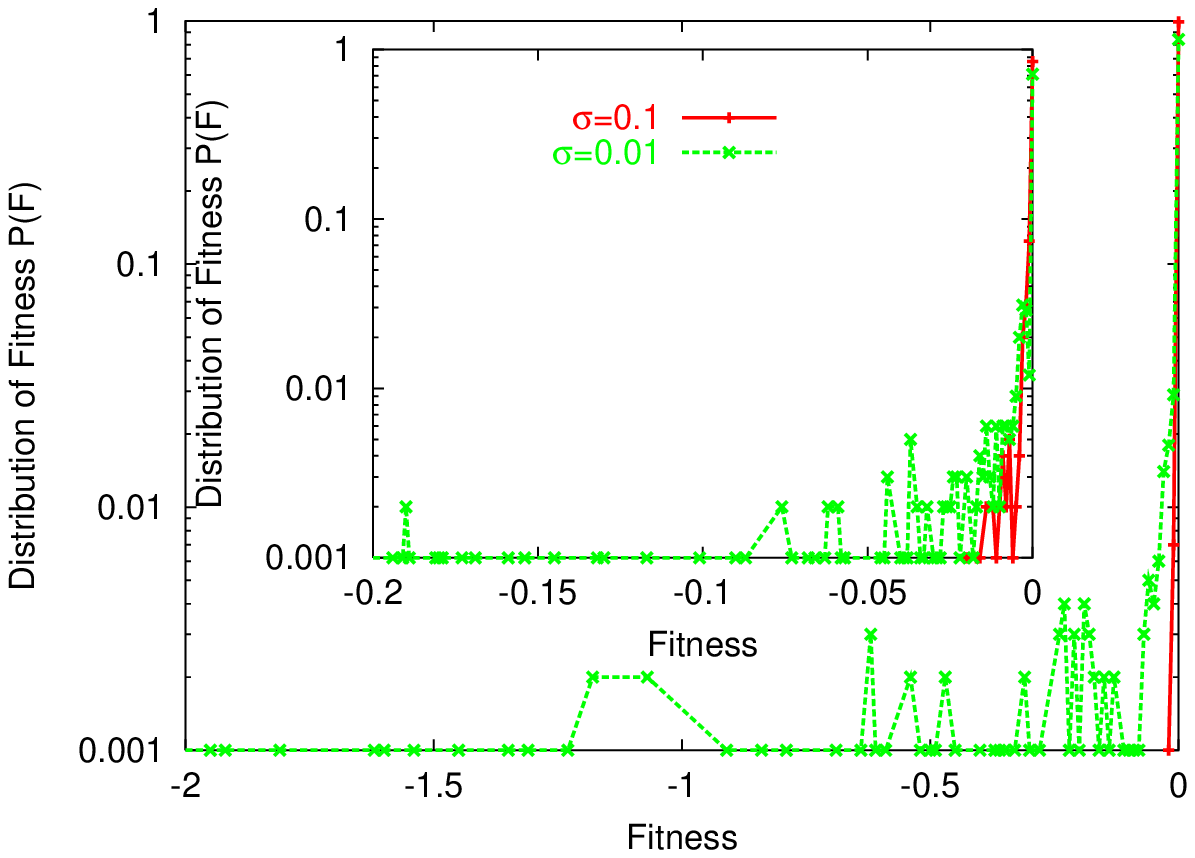}
\caption{Fitness distribution $P(\overline{F})$. From the top fitness network evolved after 200 generations, 
we generated 1000 networks by changing a single element in the $J_{i.j}s$ matrix,
and computed the average fitness $\overline{F}$ for each, to obtain the fitness distribution.  Inset is the magnification for
$-0.2<\overline{F}<0$. The histogram is computed with a bin size 0,01, whereas for the inset,
we adopt the bin size 0.001.  For high $\sigma $ (red, with $\sigma=0.1$),
the distribution is concentrated at $\overline{F}=0$, whereas for low
$\sigma$ (green, with $\sigma=0.01$), the distribution is extended to
large negative values, even after many generations.
}
\end{center}
\end{figure}

The change in the distribution between (i) and (ii) is demonstrated in Fig. 2.
There is a threshold noise $\sigma_c$, below which the distribution $P(\overline{F})$ is
broadened. As a result, the average fitness over
all individuals, $<\overline{F}>=\int \overline{F} P(\overline{F})d\overline{F}$ is low.
$<\overline{F}>$ and the lowest fitness values over individuals $\overline{F}_{min}$,
after a sufficiently large number of generations, are plotted against $\sigma$ in Figure 3.
The abrupt decrease in fitness suggests the existence of threshold noise level $\sigma_c$, below which
low-fitness mutants always remain in the distribution.

This transition on the noise level is rather sharp, as long as the number of genes $M$ is
large. As $M$ becomes smaller, the plateau in the highest fitness value against $\sigma$ is decreased, and
the transition loses sharpness, as shown in Fig. 4 for
$M=16$, with the mutation rate per single path change per generation.
With an increase in $M$, the region achieving the highest fitness with sharp distribution
increases, as well as the increase in the sharpness of transition.

If there are more genes, change in a single path has a smaller influence, so that
an increase in the robust evolution region is to be expected. However, the increase in the robust region with $M$ is not
solely caused by the decrease in the effective mutation rate. In Fig. 5, we have changed
the number of genes $M$, by fixing the mutation rate per path, i.e.,
a single path change per generation for $M=16$, 4 paths for $M=32$, and 16 for $M=64$.
Even in this fixed mutation rate, the transition is sharper, and the
increase in the robustness in the evolution is detected.

Of course, the increase in the mutation rate reduces the robust evolution region.
Comparing Fig. 2(a) and Fig. 4 shows that the threshold noise level $\sigma_c$ 
for the robust evolution region is lowered with the decrease in mutation rate, where the mutation rate
is 16 times higher for the simulation for $M=64$ in Fig. 4 than that for Fig. 2(a).

\begin{figure}[tbp]
\begin{center}
\includegraphics[width=8cm,height=6cm]{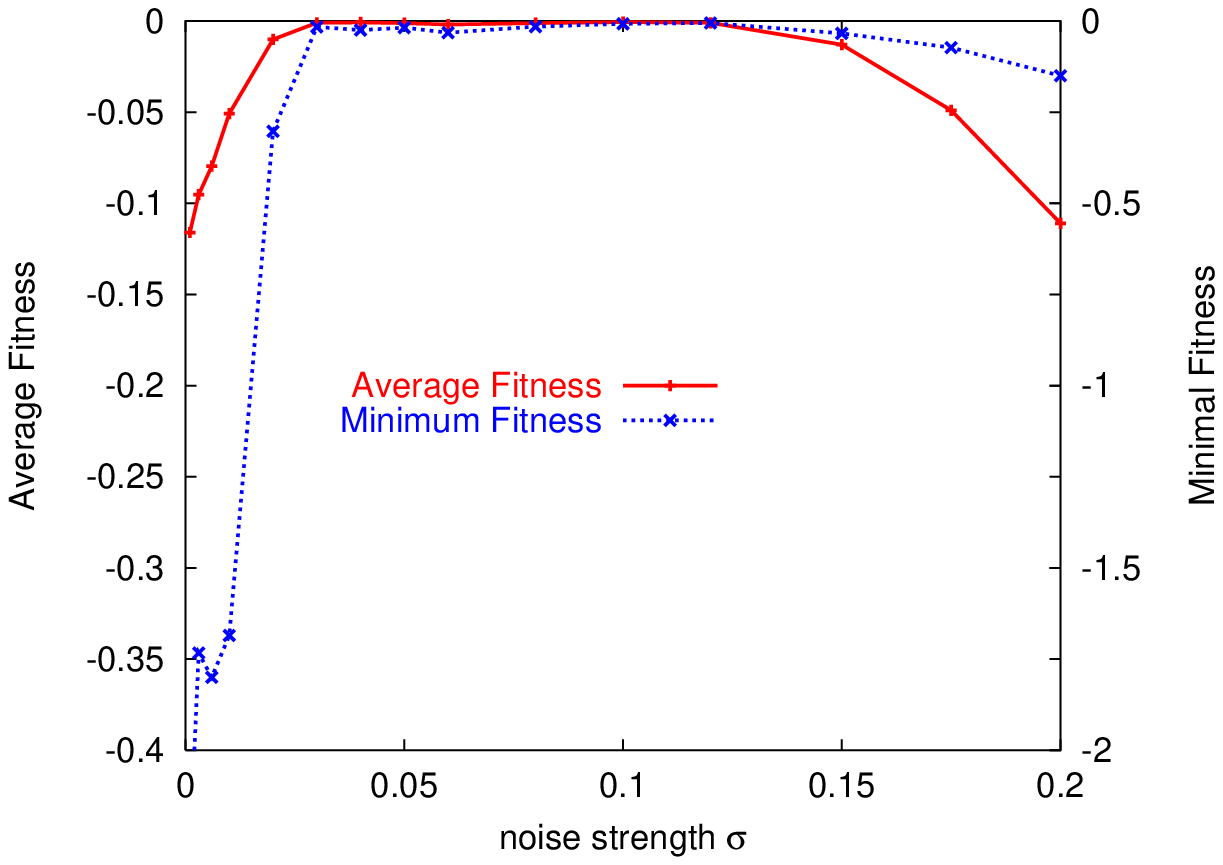}
\includegraphics[width=8cm,height=6cm]{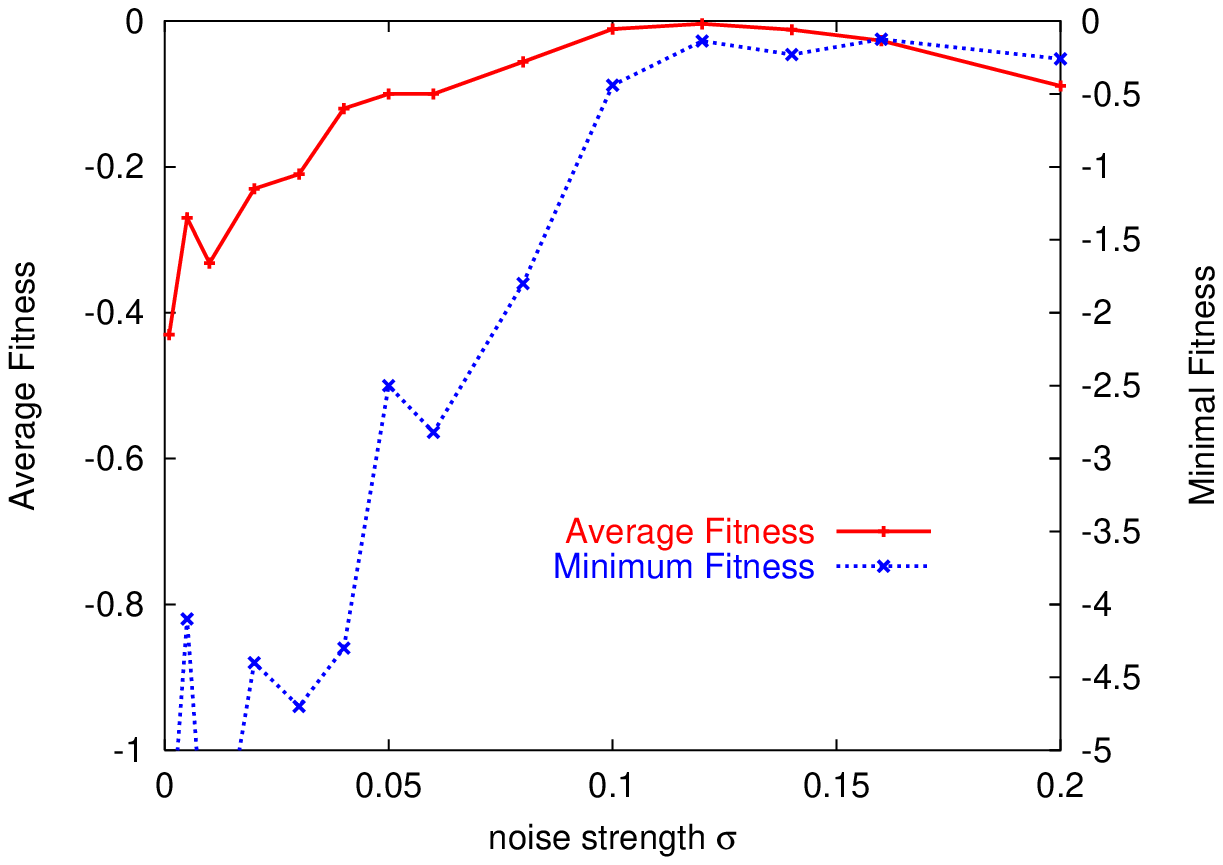}
\caption{Average of average fitness $<\overline{F}>$ and minimal fitness over generations,
plotted against the noise strength $\sigma$.
$<\overline{F}>$, the average of the average
fitness $\overline{F}$ over all individuals is computed for 100--200
generations. (For $\sigma> .1$, we choose the average over 200-300 generation,
as the fitness plateau is reached later). The minimal
fitness is computed from the time average of the least fit network present
at each generation. (a) $M=64$ (b)$M=16$.  The mutation rate is one path over all paths.
In (b), since there are 16 times paths of (a), the mutation rate for each path
is $1/16$, compared from (a).}
\end{center}
\end{figure}

\begin{figure}[tbp]
\begin{center}
\includegraphics[width=8cm,height=6cm]{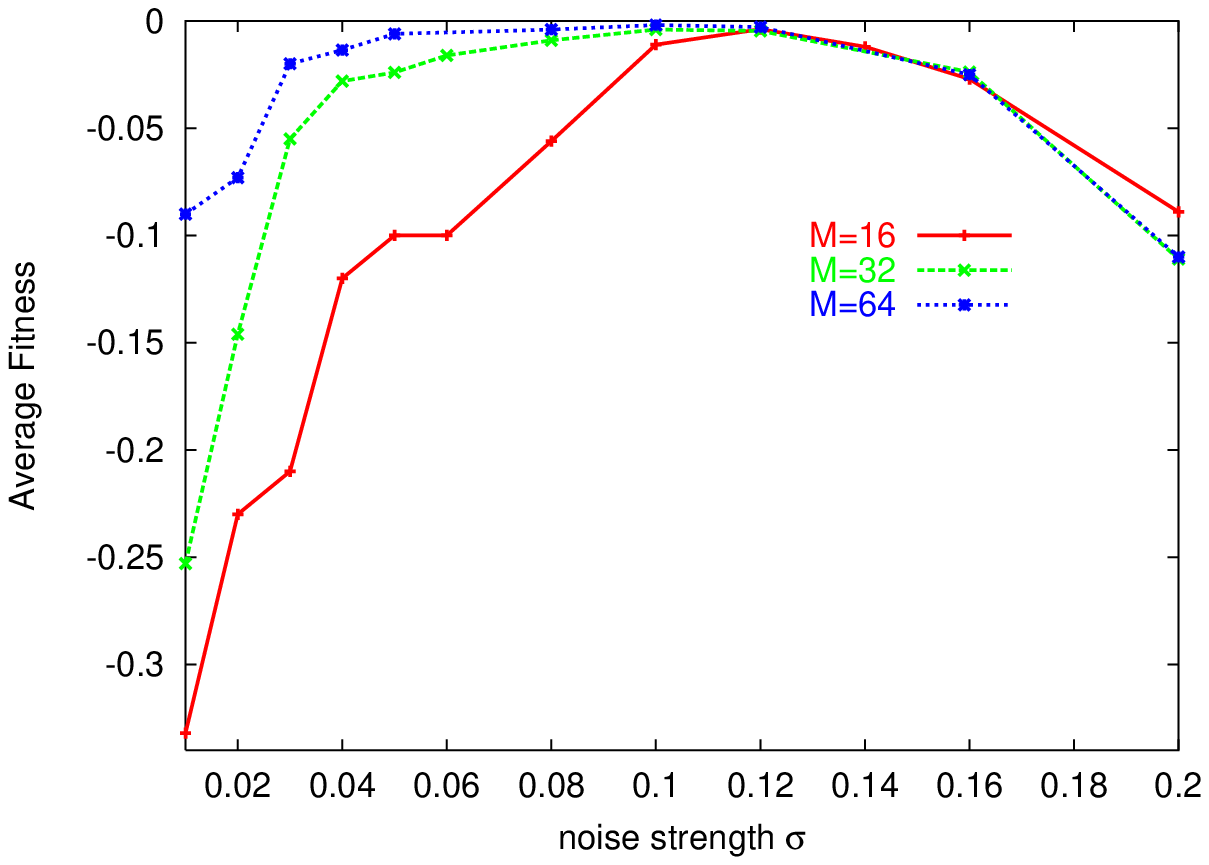}
\caption{Average of average fitness $<\overline{F}>$ over generations after transients,
plotted against the noise strength $\sigma$. $<\overline{F}>$, the average of the average
fitness $\overline{F}$ over all individuals is computed over 100--200 generations for $\sigma \leq.1$, and
200-300 generations for  $\sigma >.1$ to remove the transient effect.
The fitness for $M=16$, 32, and 64 are computed, by fixing the mutation rate at 1 per 16 x 16 paths, i.e.,
one path in total for $M=16$, four paths for $M=32$, and 16 paths for $M=64$.}
\end{center}
\end{figure}

At the transition at $\sigma \approx \sigma_c$ we discuss here,
mutational robustness of the evolved network changes. The top fitness network evolved at a low noise level ($\sigma<\sigma_c$)
does not have robustness against mutation, in contrast to those evolved at a higher noise level. To check this
distinction statistically, we measured the average fitness of mutants generated from the top fitness.
We took a network with the top fitness evolved after generations, and changed $m$ paths randomly
(i.e., change in a value of $J_{i,j}$ among $\pm 1$,0). By making a thousand networks of such $m$ mutations, we have
measured the average fitness of such mutants, which is plotted against $m$ in Fig. 6.  
For $\sigma<\sigma_c$, the average fitness decreases linearly with the number of added mutations $m$,
as $<\overline{F}>=-C(\sigma)m $. The coefficient $C(\sigma)$ decreases with the increase in $\sigma$, and
for $\sigma=.1 (> \sigma_c)$, the linear decrease component vanishes, giving rise to a plateau around $<\overline{F}>=0$. 
In other words, the fitness landscape has an almost neutral region.
The fitness is rather insensitive to mutation, demonstrating the evolution of mutational robustness.
Around $\sigma \sim \sigma _c$, $C(\sigma)$ estimated at small mutation number $m$ is rather small, and
we may expect that $C(\sigma ) \rightarrow 0$ as $\sigma \rightarrow \sigma _c$, although
careful examination of the slope at $m \rightarrow 0$ limit is needed to confirm it.

The transition with regards to the mutational robustness at $\sigma_c$ reminds us of the
error catastrophe introduced by Eigen and Schuster\cite{Eigen}, since the population of
non-fit mutants starts to accumulate after this transition.  In their error catastrophe,
the transition occurs when the advantage by fitted individuals cannot overcome 
combinatorial diversity of non-fit mutants.  Their error catastrophe occurs with the increase in
the mutation rate, while our robustness transition occurs with the decrease in developmental noise.
In our case, the transition is of dynamic origin, as will be discussed in the next section,
whereas there exist (combinatorially) much more non-fit networks than the fittest ones.  The
robustness transition here could be related with (or regarded as an extension of) error catastrophe, but
the relationship between the two remains to be clarified.

\begin{figure}[tbp]
\begin{center}
\includegraphics[width=8cm,height=6cm]{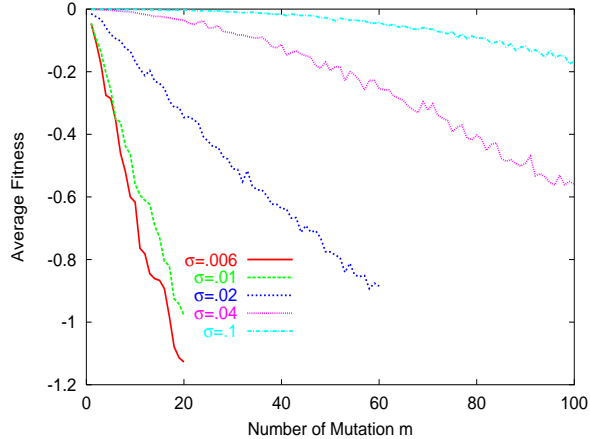}
\caption{Decline of the average fitness plotted as a function of $m$, which is the number of mutated paths
from a top-fitness network.
The average fitness is computed from $\overline{F}$'s over 1000 mutants generated from an evolved network having the top fitness,
by inserting $m$ mutations.
For $\sigma<\sigma_c$, it decreases linearly as $<\overline{F}>=-C(\sigma)m$, as a function of
the number of mutated paths $M$. Here $C(\sigma)$ decreases.  At $\sigma =.1>\sigma_c$, $C(\sigma)=0$.}
\end{center}
\end{figure}

\section{Dynamic origin of the robustness transition}

Why does the system not maintain the highest fitness state under a condition of small
phenotypic noise $\sigma<\sigma_c$? Indeed, the top fitness networks that evolved
under low noise have dynamic behaviors distinct from those that evolved under high noise.
First, the highest fitness network evolved at low $\sigma$ often fails to reach the
target if simulated under a higher noise level. The expression level often exhibits a few oscillations before
reaching the target state and noise might cause expression of the output genes to switch to `off' states.
In contrast, the temporal course of gene expression evolved for
$\sigma>\sigma_c$ is much smoother, and is not affected by noise.
This distinction is confirmed by simulating gene expression dynamics by cutting off the noise term
over a variety of initial gene expression conditions, and to check
if the orbit is attracted to the original target. We found that,
for networks evolved under $\sigma>\sigma_c$, a large portion of the initial
conditions is attracted into the target pattern, while for those evolved under
$\sigma<\sigma_c$, only a tiny fraction (i.e., the vicinity of all `off' states) is attracted to the
target (see Fig. 6).

\begin{figure}[tbp]
\begin{center}
\includegraphics[width=8cm,height=6cm]{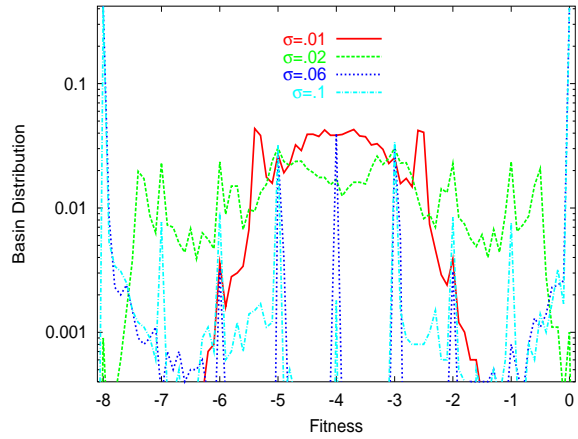}
\caption{Distribution of the fitness value when the initial condition for $x_j$
is not fixed at -1, but is distributed over $[-1,1]$. We chose the
evolved network as in Fig. 5; for each network we took 10000
initial conditions, and simulated the dynamics (1) without noise, to
measure the fitness value $F$ after the system reached an attractor (as the
temporal average $400<t<500$). The histogram is plotted with a bin size
0.1 using a semi-log plot.}
\end{center}
\end{figure}

\begin{figure}[tbp]
\begin{center}
\includegraphics[width=7cm,height=7cm]{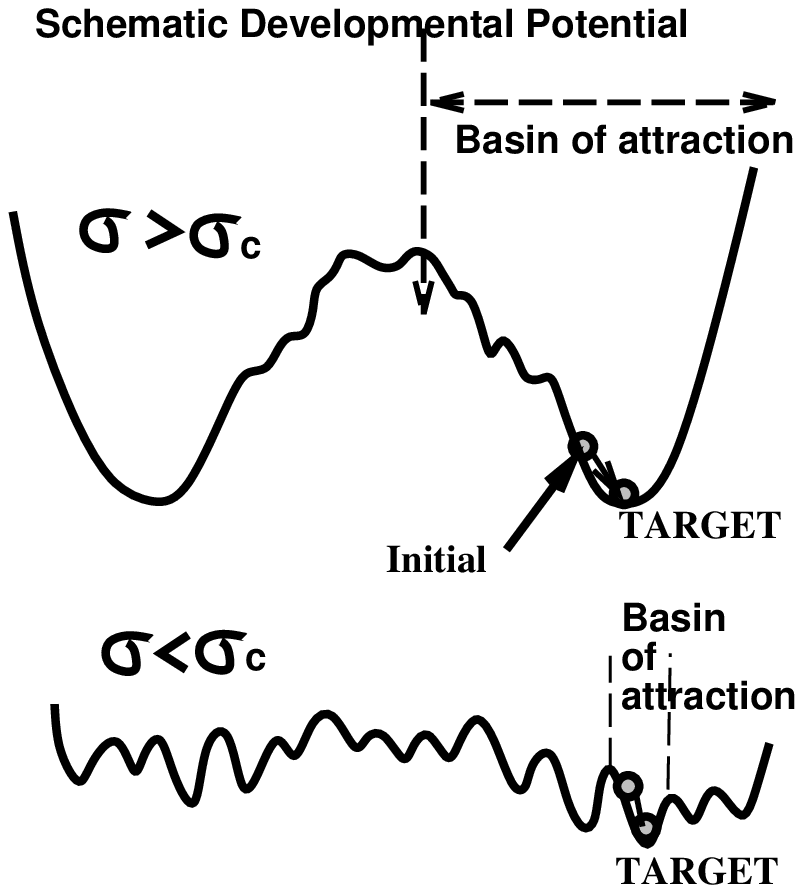}
\caption{Schematic representation of the basin structure, represented as a
process of climbing down a potential landscape. 
A smooth landscape is evolved under high noise (above), and
a rugged landscape is evolved under low noise (below).}
\end{center}
\end{figure}

If the time course of a given gene expression to reach its final pattern
is represented as a motion falling along a potential valley, our results suggest
that the potential landscape becomes smoother and simpler through evolution and
loses ruggedness after a few hundred generations.
This `developmental' landscape is displayed schematically in Fig. 7. For networks evolved under
$\sigma>\sigma_c$ there is a large, smooth attraction to the target, whereas for the dynamics evolved under $\sigma< \sigma_c$, the
initial states are split into small basins, from each of which the gene 
expression patterns reach different steady states.

Now, consider the mutation to a network. Any change in the network leads to slight alterations in gene expression dynamics. 
In smooth dynamics, as in Fig. 7(a), this perturbation influences the
attraction to the target only slightly. By contrast, under the dynamics as shown in Fig. 7(b), a slight change easily destroys the attraction to the
target attractor. For this latter case, the fitness of mutant networks is distributed down to
lower values, which explains the behavior observed in Fig. 3.

Accordingly, evolution to eliminate ruggedness in developmental potential is possible only for sufficient noise amplitude, whereas
ruggedness remains for small noise values and the developmental dynamics often
fail to reach the target, either by noise in gene expression dynamics or by mutations to the networks.
It is interesting to note that a greater set of initial conditions is attracted to a target pattern 
for networks evolved under conditions of high noise.
Existence of such global attraction in an actual gene network has recently been
reported for the yeast cell cycle\cite{Ouyang}.

\section{Evolution of robustness}

\begin{figure}[tbp]
\begin{center}
\includegraphics[width=8cm,height=6cm]{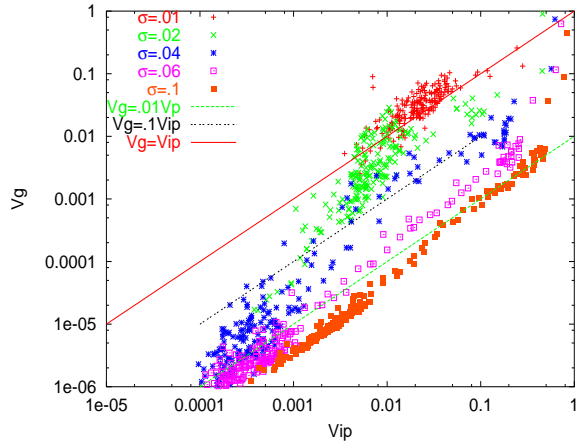}
\caption{Relationship between $V_g$ and $V_{ip}$. $V_g$ is computed from
$P(\overline{F})$ at each generation, and $V_{ip}$ by computing
the variances of isogenic fluctuation of the fitness (over $L$ runs) for existing individuals,
and averaging them.  
Plotted points are over 200 generations.
For $\sigma> \sigma_c\approx .03$, both decrease with generations. See text for the definition of
$V_{ig}$ and $V_g$.}
\end{center}
\end{figure}

According to our results in the two previous sections,
a network evolved under high noise has two types of robustness.
The target pattern is reached even if the developmental dynamics are perturbed by noise or 
by a mutation to the network. The fitness is rather insensitive to both of
such perturbations. 

As an index for robustness, variance of the fitness (or phenotype) is useful\cite{Plos1}.
There are two types of variance corresponding to the above two types of robustness, i.e.,
genetic and epigenetic robustness. Variance corresponding to genetic change,
denoted by $V_g$, is defined as phenotype variance caused by the 
distribution of genes;

\begin{equation}
V_g=\int P(\overline{F}) (\overline{F}-<\overline{F}>)^2 d\overline{F},
\end{equation}

\noindent
where we note $ P(\overline{F})$ is the distribution of average fitness over
all networks at each generation, and $<\overline{F}>= P(\overline{F})\overline{F}d\overline{F}$ is 
the average of average fitness over all networks.

As the variance decreases, the system increases its robustness to
genetic change (mutation). On the other hand, epigenetic robustness is measured by the
phenotypic variance of isogenic organisms $V_{ip}$.  Indeed, the phenotypes of isogenic individual 
organisms fluctuate, as discussed extensively\cite{Elowitz,Collins,Furusawa,Barkai,noise-review}.
We define the isogenic phenotypic variance $V_{ip}(I)$ by

\begin{equation}
V_{ip}(I)=\int p(F;I) (F-\overline{F}_I)^2 dF,
\end{equation}

\noindent
where $p(F;I)$ is the fitness distribution among isogenic individuals $I$ sharing the same network $J_{ij}$, and
$\overline{F}_I=\int F p(F;I) dF$ is the average fitness of the genotype $I$.
$V_{ip}(I)$ generally depends on the individual (genotype) $I$.  As a measure of $V_{ip}$, we adopted such $I$ that
gives the peak in the fitness distribution  $P(\overline{F})$, i.e., most typical genotype.
Or, instead, we estimated it by the average of $V_{ip}(I)$ over all genotypes $I$ existing at each generation.
The overall $V_{ip}$-$V_g$ relationship is not altered between the two estimates.  In Fig.8,
we adopted the latter estimate, to reduce statistical error.

Under high noise conditions, the selection process favors a developmental process
that is robust against it. This robustness to noise is then embedded into
robustness against mutation. Indeed, for $\sigma>\sigma_c$, both $V_{g}$ and $V_{ip}$
decrease through the course of evolution (Fig. 8), while maintaining proportionality
between the two. Such proportionality between the two has been discussed from an evolutionary stability analysis
under a few assumptions\cite{JTB,book,ECCS}.  For  $\sigma>\sigma_c$, the
the inequality $V_{ip} > V_g$ is satisfied, while it is broken at the transition  $\sigma \sim \sigma_c$,
which also agrees with the theoretical analysis.  
On the other hand, the proportionality between  $V_{ip}$ and $V_g$ is also consistent 
with an observation on an experiment in bacterial evolution \cite{Sato}, if the Fisher's theorem\cite{Fisher,Fisher2,Futuyma}
on the proportionality between evolution speed and $V_g$ is applied.

\section{Consolidation of the Expression of Non-target Genes}

\begin{figure}[tbp]
\begin{center}
\includegraphics[width=8cm,height=6cm]{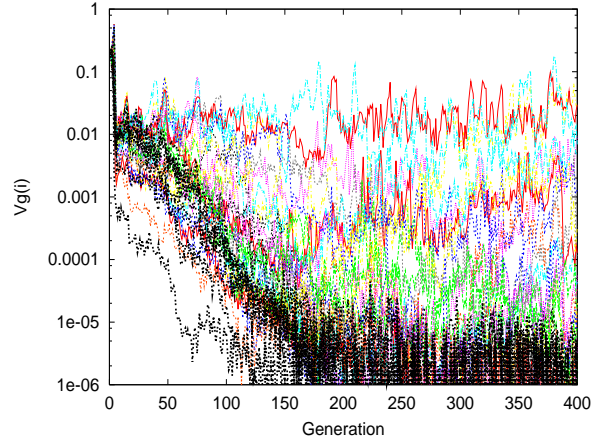}
\caption{The evolutionary change of $V_g(i)$ plotted as a function of generation.
At each generation, $V_g(i)$ is computed from
the distribution of $N=$200 individuals at each generation,
from the average value of $Sign(x_i)$ over $L=$200 runs.
The black lines are the variance for target genes $i=1,2,...,k$,
while the others are for non-target genes with $i=k+1,..,M$.
The variance values of target genes decrease at earlier generations, while
about a half of non-target genes also exhibit decreases.
$\sigma =0.1$.}
\end{center}
\end{figure}

\begin{figure}[tbp]
\begin{center}
\includegraphics[width=8cm,height=6cm]{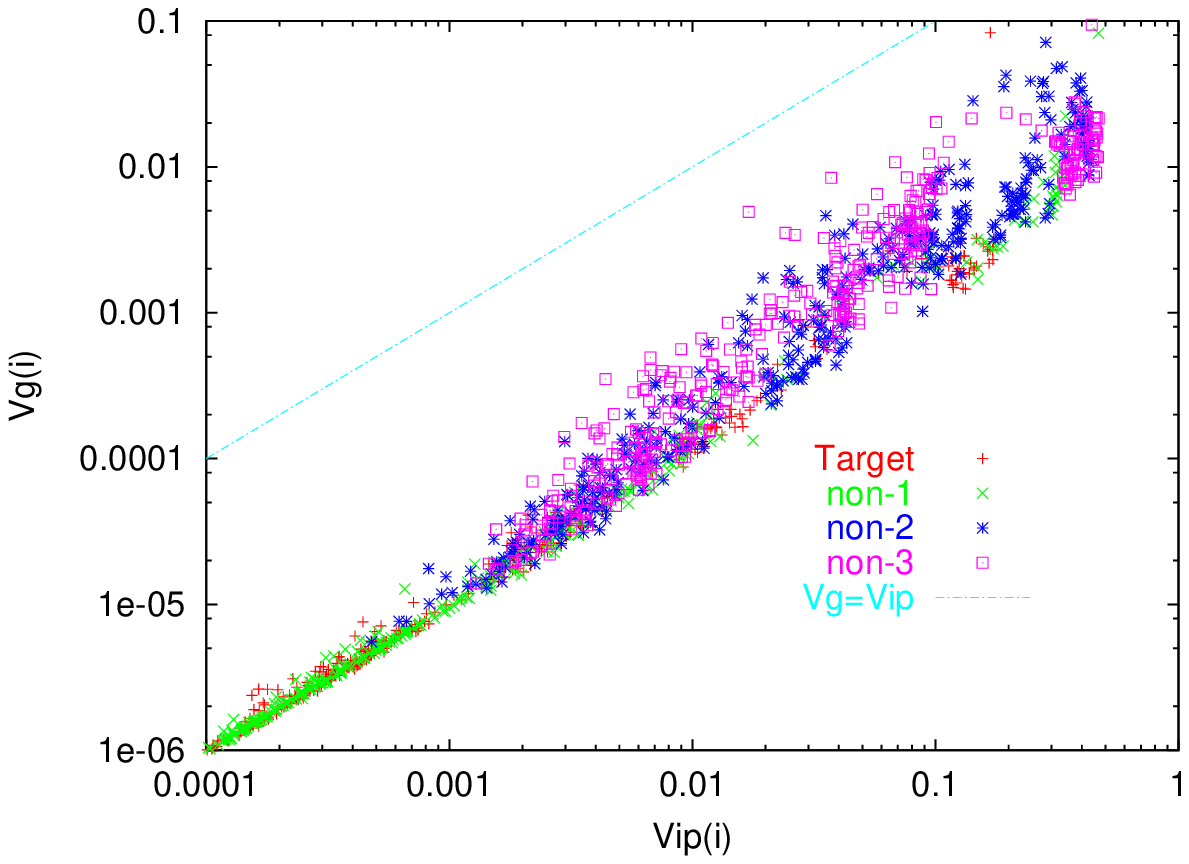}
\includegraphics[width=8cm,height=6cm]{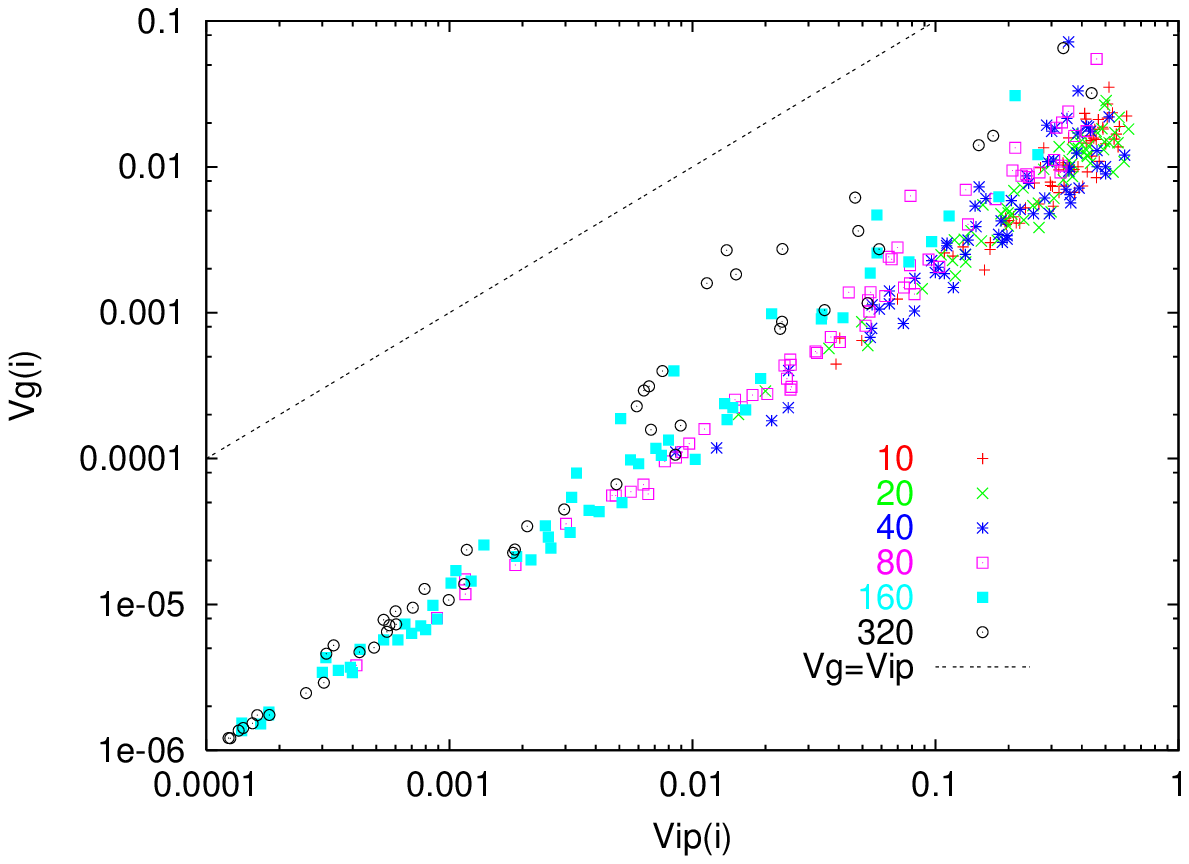}
\caption{Relationship between $V_g(i)$ and $V_{ip}(i)$. 
$V_g(i)$ is computed as a variance of the distribution of
$(\overline{Sign(x_i)})$ over 200 individuals at each generation, and $V_{ip}(i)$ as that
of the distribution of $Sign(x_i)$ over 200 runs. $\sigma=0.1$. (a) Each point is a
plot from one generation for four different genes
from one target $+$ and three non-target (x,*,$\Box$). Plotted over 400 generations,
where $(V_g(i),V_{ip}(i))$ at later generations take smaller values.
(b) The plot of $(V_g(i),V_{ip}(i))$ for all genes $i$ at the generation
10(red +), 20 (green x), 40(blue *), 80(pink $\Box$, 160(sky blue *), and 320 (black $\circ$).
}
\end{center}
\end{figure}

So far we have studied how a dynamical system consisting of positive and negative
feedbacks is shaped to generate a given target output pattern. Here, the matrix $J_{ij}$ has a large number
of nonzero elements, and generally the structure is complex. 
Fitness generated from such a complex network is robust to noise and to mutation. 
How is such a robust system achieved to fixate the expressions of target genes?

In our model, there are more degrees of freedom (genes) $x_i$ besides those for the target genes.
The expression levels of the non-target genes do not influence the fitness, and can
take either positive or negative values. Hence, there is no selection pressure leading them to be fixed,
or robust to noise or to mutation. 
However, we have found that expression levels of many (about half) non-target genes are fixed to positive or negative values
successively through the course of evolution in the robust evolution region, i.e., for $\sigma>\sigma_c$.
Even after mutation to $J_{i,j}$ values, the sign value of $x_i$ does not change for most genes $i$(data not shown).
For such genes, the variance of each expression level over mutants becomes rather small because of evolution.

To check for a possible decrease in the variance of expression levels,
we define $V_g(i)$ for gene $i$, as the variance of $\overline{Sign(x_i)}$ computed at $t=T_{ini}$, (i.e., after the time span for
development) over individuals with different genes, instead of the variance of fitness we measured in the last section.
($\overline{Sign(x_i)}$ is the average among isogenic individuals, i.e., the average over $L$ runs).
In other words, $V_g(i)$ is defined by replacing $\overline{F}$ by $\overline{Sign(x_i)}$ in eq.(3).
This variance $V_g(i)$ is plotted in Fig. 9 as a function of generation. Here, $V_g(i)$ for $i\leq k$,
the variance of the expression level of the target genes decreases at an earlier generation, together with a
few other genes, followed by a decrease in $V_g(i)$ for about half of the others towards low levels.
Indeed, after 200 generations of evolution, the variance $V_g(i)$ for about half of the genes
is less than $10^{-5}$.
Expression levels of many degrees of freedom other than the requested degrees of freedom as a target are also fixed.
This consolidation is relevant to achieve a robust system. Indeed, for $\sigma<\sigma_c$, such fixation of
non-target gene expression levels is not observed.
  
Similarly to $V_g(i)$, we define $V_{ip}(i)$ as the isogenic variance for each $Sign(x_i)$, 
by using $Sign(x_i)$ instead of $F$ in eq. (4), and have computed its evolution over generations. For 
genes which show the decrease in $V_g(i)$, the variances
$V_{ip}(i)$ also decrease through evolution. Furthermore, for such genes $i$, 
$V_g(i)$ and $V_{ip}(i)$ decrease in proportion, as shown in Fig. 10(a).
The proportionality relationship between $V_g(i)$ and $V_{ip}(i)$ holds true not only for
fitness but also for many genes, including non-target ones.
Surprisingly, as evolution progresses, the plot of 
$(V_{ip}(i),V_g(i))$ approaches a single line, suggesting that 
the proportion coefficient between the two approaches the same value for all
genes, whose gene expressions are fixated to on or off by generation. In fact, we have plotted
$(V_{ip}(i),V_g(i))$ for all genes in Fig. 10(b). For each given generation, the overlaid plot of
$(V_{ip}(i),V_g(i))$ fits on the same line for most genes. In other words,
the proportion relationship between the two variances holds not only for each gene's expression
through the course of evolution, but also for expression levels over different genes at each generation.
Over generations, the plots shift toward smaller values $(V_{ip}(i),V_g(i))$, approaching
a unique line. This `universal' coefficient for all genes
suggests the existence of a global potential dynamic system governing many gene expression patterns.

\section{Summary and Discussion}

We have shown here that a dynamical system that is robust both to noise and to structural
variation is shaped through evolution under noise. 
In our study, robustness to developmental noise and to mutation are represented quantitatively in terms of the
phenotypic variance of isogenic organisms $V_{ip}$ and by genetic variation $V_g$. 
The proportionality between the two through evolution is a quantitative manifestation
 of how developmental and mutational robustness can evolve in coordination.
In fact, whether these two types of robustness emerge under natural selection has long been 
debated in the context of developmental
dynamics and evolution theory\cite{Evolution,Wagner,Ancel-Fontana,Kirschner},
since the propositions of stabilization selection
by Schmalhausen\cite{Schmalhausen} and canalization by Waddington\cite{Waddington,Bergman}
 more than half a century ago. Here we have demonstrated a correlation between
developmental robustness to noise and genetic robustness to mutation, and have shown that
the former leads to the latter.

Isogenic phenotypic fluctuation is related to phenotypic plasticity,
which is a degree of phenotype change in a different environment.
Positive roles for phenotypic plasticity in evolution have 
been discussed elsewhere \cite{Kirschner,Eberhard,Ancel,plasticity}.
Because susceptibility to environmental changes and phenotypic fluctuation
are correlated positively according to
the fluctuation--response relationship\cite{book}, our present results
on the relationship between phenotypic fluctuations and evolution
imply a relationship between phenotypic plasticity and evolution
akin to the genetic assimilation proposed by Waddington\cite{Waddington}.

Although we have demonstrated this evolution of robustness using a
network model of transcriptional regulation,
we expect this behavior to be observable generally
if fitness is determined through developmental dynamics that are
sufficiently complex so that a given developmental process,
when deviated by noise, may fail to reach the fittest target pattern.
In fact, the decrease in phenotypic variance as well as the proportionality law
between $V_g$ and $V_{ip}$ has been confirmed in a catalytic reaction network model\cite{JTB}.

We have also found that the expression of many non-target genes is also fixed.
The behavior of many other degrees of freedom is consolidated with that directly related to fitness.
With this consolidation, the system's behavior is constrained to exhibit an identical phenotype 
against noise and mutation. 
Many degrees of freedom are highly correlated, which possibly results in the emergence of
a simple set of global dynamics over many gene expressions.
The existence of a universal proportionality between $V_g$ and $V_{ip}$ with an identical proportion coefficient over many genes 
might be a manifestation of such global dynamics.
It is interesting to note that this universal proportionality over components is
also observed in a reaction network model for a reproducing cell\cite{Furusawa2}.

During the course of evolution, the variance levels of some gene expressions decrease at an early generation,
while others decrease only later. This difference between genes should be related to how their expressions influence
those for the target. It will be important to study the order of fixation 
in terms of the network structure.

Borrowing the concept from statistical physics, we previously proposed an evolutionary fluctuation--response relationship,
where proportionality between evolution speed (response) and isogenic fluctuations
is proposed and experimentally verified. As the speed of evolution is proportional to
$V_g$ according to Fisher's fundamental theorem of natural selection\cite{Fisher,Fisher2},the relationship then
suggested the proportionality between $V_{ip}$ and $V_g$.  Note that in physics, the proportion coefficient
between response and fluctuation is represented universally in terms of temperature, and is known
as the fluctuation--dissipation relation\cite{Kubo,Vulpiani} or Einstein's relation\cite{Einstein}. Discovery of
our universal proportion coefficient over genes suggests the existence of
a general theory akin to the fluctuation--dissipation theorem that could be applied to evolutionary biology
and also to robust design of dynamical systems.

In the present paper, we adopted a simple fitness condition favoring a fixed expression of target genes.  
Accordingly, such system that has a large basin for a fixed point attractor corresponding to the target pattern is evolved.  
It should be of importance to examine a case with a more complex
fitness condition.  A straightforward extension is the use of several target patterns depending on
some inputs applied to some genes.  Another extension is the use of fitness favoring for dynamic attractors.  As the model (1), 
with suitable choice of $J_{ij}$, shows periodic or chaotic attractors\cite{Glass}, study on
the evolution of robust dynamic system is also of importance ( see also \cite{Rikvold}).

I would like to thank Chikara Furusawa, Koichi Fujimoto, Masashi Tachikawa, Shuji Ishihara,
Tetsuya Yomo, and Satoshi Sawai for stimulating discussion.

\end{document}